\documentclass[11pt]{article}
\usepackage{amsmath,amssymb,color,graphics,epsfig,cite}
\pdfoutput=1

\usepackage{amsfonts}

\newcommand{\hoch}[1]{$\, ^{#1}$}

\makeatletter
\@addtoreset{equation}{section}
\makeatother

\newcommand{\be}{\begin{equation}}
\newcommand{\ee}{\end{equation}}
\newcommand{\bea}{\setlength\arraycolsep{2pt} \begin{eqnarray}}
\newcommand{\eea}{\end{eqnarray}}
\newcommand{\nn}{\nonumber}

\def\ft#1#2{{\textstyle{\frac{\scriptstyle #1}{\scriptstyle #2} } }}
\def\fft#1#2{{\frac{#1}{#2}}}

\def\0{{\sst{(0)}}}
\def\1{{\sst{(1)}}}
\def\2{{\sst{(2)}}}
\def\3{{\sst{(3)}}}
\def\4{{\sst{(4)}}}
\def\5{{\sst{(5)}}}
\def\6{{\sst{(6)}}}
\def\7{{\sst{(7)}}}
\def\8{{\sst{(8)}}}
\def\sst#1{{\scriptscriptstyle #1}}

\def\del{{\partial}}

\begin{document}


\begin{center}

{\Large {\bf Horndeski Gravity and the Violation of Reverse Isoperimetric Inequality }}

\vspace{40pt}
{\bf Xing-Hui Feng\hoch{1}, Hai-Shan Liu\hoch{2\dag}, Wen-Tian Lu\hoch{1} and H. L\"u\hoch{1*}}

\vspace{10pt}

{\it \hoch{1}Center for Advanced Quantum Studies, Department of Physics, \\
Beijing Normal University, Beijing 100875, China}

\vspace{10pt}
\hoch{2}{\it Institute for Advanced Physics \& Mathematics,\\
Zhejiang University of Technology, Hangzhou 310023, China}

\vspace{40pt}

\underline{ABSTRACT}
\end{center}

We consider Einstein-Horndeski-Maxwell gravity, together with a cosmological constant and multiple Horndeski axions. We construct charged AdS planar black holes in general dimensions where the Horndeski anxions span over the planar directions.  We analyse the thermodynamics and obtain the black hole volumes.  We show that the reverse isoperimetric inequality can be violated, implying that these black holes can store information more efficiently than the Schwarzschild black hole.

\vfill {\footnotesize Emails: \hoch{\dag}hsliu.zju@gmail.com\ \ \ \hoch{*}mrhonglu@gmail.com}

\thispagestyle{empty}

\pagebreak

\tableofcontents
\addtocontents{toc}{\protect\setcounter{tocdepth}{2}}


\newpage

\section{Introduction}

There has been tremendous progress in understanding the properties of black holes, classically much more than quantum mechanically.  There is however a lack of understanding of one rather mundane concept, the volume of a black hole.  The black hole thermodynamics appears to imply the membrane paradigm \cite{Thorne:1986iy} or more generally the holographic principle \cite{Susskind:1994vu}, which makes the black hole event horizon more physically relevant. Nevertheless, a black hole does occupy the space and we would like to know its volume.  A naive proposal is to simply integrate out the volume density over the ``space'' encircled by the horizon, namely
\be
V=\int_{\rm inside} d^{D-1} x\, \sqrt{-g}\,.
\ee
For the Schwarzschild black hole in four dimensions, this gives precisely $\ft43\pi r_0^3$, which is the same as the Euclidean volume of a spherical ball of radius $r_0$.  This definition however cannot withstand close scrutiny. Firstly, the time coordinate $t$ is not globally defined crossing over the horizon, and the constant-$t$ slice inside the horizon is not necessarily spacelike. Secondly, the metric inside the horizon may not be uniquely defined even when the spacetime outside the horizon is completely fixed, owing to the locality of Einstein's equations. For example, the point-source Schwarzschild black hole has the same spacetime metric outside the horizon as that of black hole generated by a spherical shell of the same mass.  Integrating the spacetime geometry inside the horizon can obviously lead to different answers even though the black holes look identical from the outside of the horizon.

For asymptotic anti-de Sitter (AdS) black holes, the volumes can be determined indirectly by means of the first law of black hole thermodynamics. The negative cosmological constant $\Lambda$ contributes to the energy-momentum tensor like perfect fluid with positive pressure $P=-\Lambda/(8\pi)$. Treating this pressure as a thermodynamic variable, one can derive its conjugate thermodynamic volume $V_{\rm th}$ by completing the black hole first law
\be
dM=T dS + V_{\rm th} dP + \cdots\,.
\ee
With the inclusion of $(V_{\rm th}, P)$, the black hole mass $M$ should then be interpreted as the enthalpy of the thermodynamic system \cite{Kastor:2009wy,Cvetic:2010jb}.  Interestingly, the volume obtained by this means for the Schwarzschild AdS black hole turns out to be precisely the same as the naively-expected Euclidean volume.  Since the volume is independent of the cosmological constant, it is expected to be true even for the usual asymptotically-flat Schwarzschild black hole.

For a generic AdS black hole, it requires an exact solution in order to determine its volume by means of completing the first law.  By contrast, the entropy can be determined locally as $1/4$ of the area of the horizon in Einstein gravity, or more generally by the Wald entropy formula \cite{wald1,wald2}, which requires only the near horizon geometry.  An integral form of calculating the volume can also be derived from the Wald formalism \cite{Urano:2009xn,Couch:2016exn}, which necessarily requires exact solutions. Interestingly the integration is more naturally performed from horizon to asymptotic infinity \cite{Couch:2016exn}.  Recently, a local formula of determining the volume of a black hole was proposed for static black holes, and was tested against a large number of known black holes \cite{Feng:2017wvc}.

     Having obtained both the entropy (or area) and volume, it is of interest to compare the two quantities. In Euclidean space, for a fixed enclosing area, the spherical ball has the largest volume.  For black holes, it appears to be opposite. By examining large number of black holes, static or rotating ones, Cveti\v c, Gibbons, Kubiz\v n\'ak and Pope noticed that thermodynamic volumes appear to be equal or more than the corresponding Euclidean volumes associated with the entropy \cite{Cvetic:2010jb}. Physically, it implies that for a given entropy, the Schwarzschild black hole occupies the least volume, and hence it is most efficient in storing information.  This reverse isoperimetric inequality (RII) was conjectured to be generally true for all black holes \cite{Cvetic:2010jb}.

    The CGKP conjecture is rather robust.  In fact for the local volume formula of static black holes proposed in \cite{Feng:2017wvc}, the CGKP conjecture becomes a simple consequence of the null-energy condition.  However, counter examples were also found in some super rotating solutions where the horizon geometry becomes no longer compact \cite{Hennigar:2014cfa,Hennigar:2015cja}.  However, no counter example was found for static black holes in Einstein gravities.

Since the CGKP conjecture may be related to the energy condition for black holes with compact horizons \cite{Feng:2017wvc}, we expect that the RII will be violated in higher derivative gravities, which typically involve ghost excitations.  A simple example is provided by Einstein gravity with a cosmological constant $\Lambda$, extended with quadratic curvature invariants $\alpha R^2 + \beta R^{\mu\nu} R_{\mu\nu}$.  The theory contains inevitable ghost-like massive spin-2 modes \cite{Stelle:1976gc,Stelle:1977ry}, except at some critical point where the massive spin-2 modes become log modes \cite{Lu:2011zk,Deser:2011xc}. In this theory, the Schwarzschild AdS black hole with $f=-\fft13\Lambda r^2 +1 -2\mu/r$ remains to be an exact solution.  (Recently, further new black holes was numerically established to exist \cite{Lu:2015cqa}.) The mass \cite{Deser:2002rt} and entropy \cite{Lu:2011zk} however are modified to become
\be
M=(1+2\Lambda\gamma)\mu\,,\qquad S=(1+2\Lambda\gamma) \pi r_0^2\,,\label{ms0}
\ee
where $\gamma =4\alpha + \beta$.  Completion of the first law implies
\be
V_{\rm th} = \ft43\pi (1+ \gamma \Lambda) r_0^3 -4\pi\gamma r_0\,.\label{vth0}
\ee
It can be easily checked that the RII can be indeed violated.  Recently, the violation was also observed \cite{Hennigar:2016gkm} in black holes of Einsteinian cubic gravity \cite{Bueno:2016xff,Bueno:2016lrh}.  Although the theory is engineered to be ghost free in the AdS vacuum, the violation might be indicative that ghost excitations may arise in the black hole background.

As we saw above, a new subtlety emerged for higher derivative gravities.  The entropy based on the Wald entropy formula is no longer purely geometrical.  We can thus define the Euclidean volume
associated with a black hole based either on its entropy $S$, or on its geometric area ${\cal A}$ of the horizon, namely
\be
V_{S}\equiv \fft{\omega}{D-1} \big(\fft{4S}{\omega}\big)^{\fft{D-1}{D-2}}\,,\qquad
\hbox{or} \qquad V_{{\cal A}}\equiv \fft{\omega}{D-1} \big(\fft{\cal A}{\omega}\big)^{\fft{D-1}{D-2}}\,.
\ee
These two definitions are identical in Einstein gravity, but different when this theory has higher-order extensions. The RII can thus have two statements
\be
\hbox{(i):}\qquad V_{\rm th}\ge V_{S}\,,\qquad \hbox{or}\qquad
\hbox{(ii):}\qquad V_{\rm th}\ge V_{\cal A}\,.\label{tworii}
\ee
Note that the RII for black hole with (\ref{ms0}) and (\ref{vth0}) in above quadratically-extended gravity can be violated for the both statements.

   We are interested in examining the RII conjecture in ghost-free higher-derivative gravities. The simplest example may be the Einstein-Gauss-Bonnet theory.  The AdS black hole was constructed in \cite{Boulware:1985wk,Cai:2001dz}. The entropy and volume can be easily obtained, given by
\be
S=\ft14\omega r_0^{D-2} \Big(1 + 2(D-2)(D-3)k\fft{\alpha}{r_0^2}\Big)\,,\qquad
V_{\rm th}=\fft{\omega }{D-1} r_0^{D-1}\,,\label{gbsv}
\ee
where $\alpha$ is the Gauss-Bonnet coupling constant and $k=1,0,-1$ is the topological parameter.
Thus the RII of statement (ii) in (\ref{tworii}) is saturated, but the RII of (i) will be violated for positive $\alpha$, for spherically-symmetric black holes with $k=1$.

      In this paper, we consider Einstein-Horndeski-Maxwell gravity \cite{Horndeski:1974wa} with multiple Horndeski axions.  In this theory, axion fields $\chi_i$'s couple non-minimally to the Einstein tensor, namely $G_{\mu\nu}\partial^\mu \chi_i \partial^\nu\chi_i$.  Although the theory involves four derivatives; however, each field at the equations of motion has at most two derivatives acting on directly.  It follows that the linearized theory is of two derivatives, analogous to Einstein-Gauss-Bonnet gravity.  This implies that the theory can be ghost free, provided that the kinetic term of each mode is non-negative.

     We construct new AdS planar black holes in general dimensions where the Horndeski axions span over the planar directions. We analyse the global structure and obtain the thermodynamic volumes by establishing the first law of black hole thermodynamics.  Comparing the entropies and volumes, we find that the RII can be violated in both statements of (\ref{tworii}), even within the no-ghost condition.

     The paper is organized as follows.  In section 2, we present the Einstein-Horndeski-Maxwell gravity with multiple Horndeski axions. We consider the ansatz for charged AdS planar black holes and derive the equations of motion. We determine the first law of black hole thermodynamics up to the point where only the mass and volume are to be determined.  We also provide the no-ghost condition on the parameters.  In section 3, we present exact solutions of charged AdS planar black holes in four and five dimensions.  We determine all the thermodynamic quantities including the volume, by completing the first law.  In section 4, we construct the solutions to general dimensions and study their properties.  In section 5, we discuss the RII and show that it can be violated by our black holes.  We conclude the paper in section 6.

\section{Horndeski gravity with multiple axions}

\label{sec:horndeski}

\subsection{The theory and covariant equations of motion}

\label{sec:theoryeom}

In this paper, we consider Einstein-Horndeski gravity \cite{Horndeski:1974wa} with $N$ number of Horndeski axions $\chi_i$. We also include a negative cosmological constant $\Lambda<0$ and  a Maxwell field $A$ with its field strength $F=dA$.  The full Lagrangian in general $D$ dimensions is given by
\be
\fft{\cal L}{\sqrt{-g}} = \kappa(R-2\Lambda - \ft14 F^2)-
\ft{1}{2}(\alpha g^{\mu\nu}-\gamma G^{\mu\nu})\, \sum_{i=1}^N
 \del_\mu\chi_i\, \del_\nu\chi_i \,.\label{lag3}
\ee
The corresponding covariant equations of motion from the variation of the metric, axion $\chi_i$'s and $A_\mu$ are respectively given by
\bea
&& \kappa (G_{\mu\nu} +\Lambda g_{\mu\nu} - \fft12 F_{\mu\nu}^2 + \fft18 F^2 g_{\mu\nu})
 -\sum_i^{2}\ft12\alpha \Big(\partial_\mu \chi_i \partial_\nu \chi_i - \ft12 g_{\mu\nu} (\partial\chi_i)^2\Big) \cr
 &&- \sum_{i=1}^{N} \ft12\gamma \Big(\ft12\partial_\mu\chi_i \partial_\nu \chi_i R - 2\partial_\rho
\chi_i\, \partial_{(\mu}\chi_i\, R_{\nu)}{}^\rho
- \partial_\rho\chi_i\partial_\sigma\chi_i\, R_{\mu}{}^\rho{}_\nu{}^\sigma  \cr &&
-(\nabla_\mu\nabla^\rho\chi_i)(\nabla_\nu\nabla_\rho\chi_i)+(\nabla_\mu\nabla_\nu\chi_i)
\Box\chi_i + \ft12 G_{\mu\nu} (\partial\chi_i)^2\cr
&&-g_{\mu\nu}\big[-\ft12(\nabla^\rho\nabla^\sigma\chi_i)
(\nabla_\rho\nabla_\sigma\chi_i) + \ft12(\Box\chi_i)^2 -
  \partial_\rho\chi_i\partial_\sigma\chi_i\,R^{\rho\sigma}\big]\Big)  = 0
\,,\nn\\
&&\nabla_\mu \big( (\alpha g^{\mu\nu} - \gamma G^{\mu\nu}) \nabla_\nu\chi_i \big) = 0\,, \qquad
 \nabla_\nu F^{\nu\mu}  =  0\,.\label{generaleom}
\eea
Although Horndeski gravity involves a total of four derivatives, each field in the equations of motion above has at most two derivatives acting on directly.  Consequently, the linearized theory in any background contains only two derivatives.  Ghost excitations typically associated with higher-derivative theories can be avoided.  However conditions on parameters still have to be imposed so that the kinetic terms of the linearized modes are non-negative.  Without loss of much generality, we shall set $\kappa=1$ throughout of the paper, with the understanding that an overall $1/(16\pi)$ factor in the action.

Since we consider the case with a negative cosmological constant $\Lambda$, the theories all admit the AdS vacuum. In this vacuum, the effective kinetic term of the Horndeski axions becomes
\be
L_{(\chi_i, \rm kin)}= \ft12 (\alpha + \gamma \Lambda) (-g^{00})  \sum_{i=1}^N  \dot\chi^2\,.
\ee
The ghost-free condition requires that \cite{Jiang:2017imk}
\be
\alpha + \gamma \Lambda \ge 0\,\qquad {i.e.}\qquad \gamma \le \fft{\alpha}{(-\Lambda)}\,.\label{noghost}
\ee
In this paper, we assume that the constant $\alpha$ is positive, and hence any negative $\gamma$ satisfies the no-ghost condition. The no-ghost condition also allows a small window of positive $\gamma$ values. (The theory reduces to Einstein theory with minimally-coupled matter when $\gamma=0$.)  There is a critical situation when the parameters satisfy
\be
\alpha +\gamma \Lambda=0\,,\label{critical}
\ee
for which case the kinetic terms of the axions vanish, analogous to the critical point in Einstein-Guass-Bonnet gravity discussed in \cite{Fan:2016zfs}.

\subsection{AdS planar ansatz}

\label{sec:adsansatz}

In this paper, we construct electrically-charged AdS planar black holes.  It is clear that Reissner-Nordstr\o m (RN) black holes with all $\chi_i=0$ are solutions.  The RII can be easily shown to be saturated by these solutions \cite{Cvetic:2010jb}.  In this paper, we are thus interested in black holes with non-vanishing $\chi_i$.

For $\Lambda=0$, a no-go theorem was established that there is no asymptotically-flat neutral black holes involving the axion field \cite{huinic,sotzho1}.  This no-go theorem can be easily overcome by considering charged black holes \cite{ac}.  With $\Lambda\ne 0$, black hole solutions with a single axion that depends on the radial coordinate (or even time) were constructed in \cite{aco,Rinaldi:2012vy,Babichev:2013cya,ac}.  Thermodynamics of such black hole were analysed in \cite{Feng:2015oea,Feng:2015wvb}, where it was shown that the Wald entropy formula is no longer valid owing to the branch-cut divergence of the axion on the horizon.

When the theory involves multiple axions, one can consider ``magnetic'' axion ansatz such that the axions span over the planar directions.  Such a four-dimensional solution was first constructed in \cite{Jiang:2017imk}.  Holographic transport properties with momentum dissipation by the Horndeski axions were studied in \cite{Jiang:2017imk,Baggioli:2017ojd}, focusing the properties at the critical region (\ref{critical}). Here we generalize the construction to general $D$ dimensions.  We consider
\bea
ds^2_D &=& -h(r) dt^2 + \fft{dr^2}{f(r)} + r^2 (dx_1^2 + \cdots + dx_{D-2}^2)\,,\cr
A &=& a(r)\, dt\,,\qquad \chi_i=
\left\{
  \begin{array}{ll}
    \lambda x_i, &\qquad i=1,2,\ldots,D-2\,, \\
    0, &\qquad i\ge D-1\,.
  \end{array}
\right.
\eea
Note that we assume that there is a sufficient number of axions, i.e.~$N\ge D-2$. With this ansatz, the scalar equations of axions $\chi_i$ are automatically satisfied and $\lambda$ can be viewed as an integration constant. The Maxwell equation implies
\be
a'=\fft{Q}{r^{D-2}}\,\sqrt{\fft{h}{f}}\,,\label{aeom}
\ee
where $Q$ is an integration constant, parameterizing the strength of the electric charge.

The Einstein equations reduce to the following two first-order differential equations of the metric functions $(h,f)$
\bea
&&\fft{\sigma f'}{f}  + \fft{2\Lambda r}{(D-2) f} + \fft{Q^2}{2(D-2)r^{2D-5} f}
+\fft{2(D-3)f + \alpha\lambda^2}{2rf} - \fft{(D^2-9D+22) \gamma\lambda^2}{4r^3}=0\,,\nn\\
&&\qquad\qquad\qquad\qquad \sigma \Big(\fft{f'}{f} - \fft{h'}{h}\Big)=\fft{\gamma\lambda^2}{r^3}\,,\qquad
\sigma\equiv 1 - \fft{(D-4)\gamma\lambda^2}{4r^2}\,.\label{hfeom}
\eea
It can be easily shown that $h$ can be expressed as
\be
h = \sigma^{-\fft{2}{D-4}} f\,.
\ee
We can thus solve (\ref{aeom}) for the electric potential $a(r)$ and obtain
\be
a=-\fft{Q}{(D-3) r^{D-3}}\, _2F_1[\ft{1}{D-4}, \ft12(D-3);
\ft12(D-1); \ft{(D-4)\gamma\lambda^2}{4r^2}]\,.
\ee
We have chosen the gauge such that $a$ vanishes at the asymptotic infinity.  We have thus only one equation and one function left to solve, namely the $f$ equation in (\ref{hfeom}). We shall present and study the solutions in sections \ref{sec:d4d5} and \ref{sec:gend}.

\subsection{The thermodynamics of black holes}

\label{sec:thermal}

The goal of this paper is to construct AdS planar black holes and using the first law of thermodynamics to compute their volumes.  Before obtaining exact solutions, many properties can be established using the equations of motion only.  Assuming that a black hole exists with the event horizon located at $r=r_0$ for which $h(r_0)=0=f(r_0)$, we can determine its temperature
\be
T = \fft{\sqrt{h'f'}}{4\pi}\Big|_{r=r_0}=
\fft{-4\Lambda r_0^2 -(D-2)\alpha\lambda^2 - Q^2r_0^{-2(D-3)}}{8\pi (D-2) r_0\,\sigma_0^{(D-3)/(D-4)}}\,.\label{temp}
\ee
Here, we define $\sigma_0\equiv \sigma(r_0)$. It is clear that a well-defined event horizon must satisfy
\be
\sigma_0=1-\fft{(D-4)\gamma\lambda^2}{4r_0^2}>0\,.\label{r0cons}
\ee
Interestingly, this condition is always satisfied for negative $\gamma$, which also ensures the ghost-free condition.  We employ the Wald entropy formula to calculate the entropy of the black holes, given by
\be
S=\ft14 \omega r_0^{D-2} \Big(1 - \fft{(D-2)\gamma\lambda^2}{4r_0^2} \Big)\,.\label{genentropy}
\ee
Here we assume that the volume of the space $dx^i dx^i$ in the planar direction is $\omega$.
The black hole in a well-defined theory with no ghost excitations should not have negative entropy.  When $\gamma$ is positive, there is however a potential danger of having negative entropy.  To study this, we note that the solution has the extremal limit of zero temperature.  Consequently the horizon radius has a minimal value for a given $\lambda$; it is given by
\be
r_0^2\ge r_{\rm min}^2  = \fft{(D-2)\alpha \lambda^2}{4(-\Lambda)}\,.\label{minradius}
\ee
Note that the above $r_{\rm min}$ is the radius of the extremal black hole with $Q=0$.  For non-vanishing $Q$, the extremal radius becomes larger. Thus the minimal value of the entropy of the black holes for a given $\lambda$ is
\be
S_{\rm min} = \fft{\omega(D-2)}{16(-\Lambda)}\lambda^2 \big(\alpha + \gamma \Lambda\big)r_0^{D-4}
\ee
The entropy is therefore non-negative provided that the no-ghost condition (\ref{noghost}) is satisfied.
Note also that if $r_0=r_{\rm min}$ satisfies (\ref{r0cons}), then the minimum black hole radius can be reached; otherwise, the solution becomes singular before the horizon shrinks to $r_{\rm min}$.

Analogously, even before we construct the full black hole solution, we can determine the electric charge $Q_e$ and its thermodynamic conjugate $\Phi_e$, namely
\be
Q_e = \fft{1}{16\pi} \int {*F} = \fft{\omega}{16\pi} Q\,,\qquad
\Phi_e = a(\infty) - a(r_0)=-a(r_0)\,.
\ee

Typically, coupling constants in a theory should not be treated as a thermodynamic variable. However, the cosmological constant in AdS is special in that it can be viewed as an integration constant of an equivalent $D$-form field strength of some $(D-1)$-form gauge potential.  The cosmological constant can then be treated as a variable, namely the thermodynamic pressure $P$.  The first law of black hole thermodynamics becomes
\be
dM=TdS + \Phi_e dQ_e + V_{\rm th} dP\,,\label{fl}
\ee
where $M$ is the mass of the black hole, and should be interpreted as enthalpy now \cite{Kastor:2009wy,Cvetic:2010jb}.  The quantity $V_{\rm th}$ is the thermodynamic volume conjugate to the pressure.  The determination of the mass and volume relies on solving the $f$ equation in (\ref{hfeom}).  We shall carry out this task in sections 3 and 4 for various dimensions.

Before finishing this subsection, we point out that in the above discussion, we treated the integration constant $\lambda$ as a fixed thermodynamic variable, i.e.~it does not vary thermodynamically.  The physical interpretation of $\lambda$ is the ``magnetic'' charge of the axions, namely
\be
\Gamma^i = \fft{1}{16\pi} \int d\chi_i = \fft{L}{16\pi} \lambda\,.
\ee
For simplicity, we assume that $\int dx_i=L$ and hence $\omega =L^{D-2}$. It follows from the axion equations that its potential is given by
\be
u_i(r) =\nu_i(r)-\nu_i(r_0)\,,\qquad \nu_i(r)= \lambda L^{D-3}\int^r dr\, \sqrt{-g} (\alpha g^{ii} - \gamma G^{ii})\,.\label{nuint}
\ee
where we have chosen the gauge such that $u_i$ vanishes on the horizon. Unlike the electric potential of Maxwell field, for which the difference between the horizon and asymptotic infinity is finite.  The above value diverges asymptotically.  We thus need to renormalize the magnetic potential to a finite value $\nu_i (r_0)$, up to an additive numerical constant.  The first law is then modified to become
\be
dM=TdS + \Phi_e dQ_e + V_{\rm th} dP + \nu_i d\Gamma^i\,.\label{fl2}
\ee
Owing to the complication arising from the inclusion of this thermodynamic quantity, we consider in this paper the simplified situation where $\Gamma^i$'s are fixed thermodynamically.  We shall come back to this point briefly in section \ref{sec:d4d5}.

\subsection{Ghost-free condition}

\label{sec:noghost}

In subsection \ref{sec:theoryeom}, we established that the theory is ghost free in the AdS vacuum provided that the parameter $\gamma$ for the Horndeski term satisfies (\ref{noghost}).  In this section, we would like to examine the same condition in the AdS planar black holes.  The kinetic term of the axion perturbations take the form
\be
L_{(\delta\chi_i,{\rm kin})}=\ft12 K(r) (-g^{00}) \delta\dot \chi_i \delta \dot \chi_i\,,\qquad
K(r)=\alpha - (D-2)\fft{\gamma f'}{2r} - (D-2)(D-3)\fft{\gamma f}{2r^2}\,.
\ee
The ghost-free condition requires that $K$ is non-negative on and outer of the event horizon.  Asymptotically, we have $f\sim -2\Lambda r^2/((D-1)(D-2))$, and hence we recover the condition (\ref{noghost}).  On the horizon, making use of the equations of motion, we find
\be
K(r_0)=\left(\alpha +\gamma  \Lambda +\fft{\gamma  Q^2}{4 r_0^{2(D-2)}}+\frac{\alpha  \gamma  \lambda ^2}{2 r_0^2}\right) \sigma(r_0)\,.
\ee
For positive $\gamma$, this quantity is clearly non-negative provided that the ghost-free condition (\ref{noghost}) is satisfied.  For negative $\gamma$, it is advantageous to re-express $K(r_0)$ in terms of temperature defined in the previous subsection. We find
\be
K(r_0)= \alpha - 2(D-2)\pi\gamma \sigma_0^{\fft{1}{D-4}}\, \fft{T}{r_0}\,.
\ee
It is then clear that for negative $\gamma$, we have $K(r_0)>0$ also for black holes.  The above calculation is indicative that the perturbations of $\chi_i$'s do not give rise to ghost modes. Of course, one needs to establish that $K(r)$ is non-negative for all the $r\ge r_0$, which can be easily done once the analytical solution for $f$ is obtained in the next section.  The ghost-free condition in four dimensions was also studied in \cite{Jiang:2017imk}.

\section{Charged AdS planar black holes $D=4,5$}

\label{sec:d4d5}

    In this section, we solve the function $f$ in (\ref{hfeom}) in four and five dimensions.
The resulting solutions describe charged AdS planar black holes.  We analyse the global structure and obtain the first law of thermodynamics.  We focus on the discussion of the black hole volumes.

\subsection{Four dimensions}

In four dimensions, the equations discussed in the previous section become degenerated, and hence it should be solved on its own.  The solutions were first constructed in \cite{Jiang:2017imk} and it is given by
\bea
ds^2 &=& e^{\fft{\gamma\lambda^2}{4r^2}}\,\big(-\tilde f dt^2 + \fft{dr^2}{\tilde f}\big) +
r^2 (dx_1^2 + dx_2^2)\,,\cr
A &=& \frac{\sqrt{\pi } Q \text{erfi}\left(\frac{\sqrt{\gamma } \lambda }{2 r}\right)}{\sqrt{\gamma } \lambda }\, dt\,,\qquad \chi_i=\lambda x_i\,,
\eea
where the function $\tilde f$ is given by
\be
\tilde f=e^{\frac{\gamma  \lambda ^2}{4 r^2}} \left(-\ft16\lambda ^2 (3 \alpha +\gamma  \Lambda )-\ft13 \Lambda  r^2\right)-\frac{\mu }{r} +
\frac{\pi \text{erfi}\left(\frac{\sqrt{\gamma } \lambda }{2 r}\right) }{12 \sqrt{\gamma } \lambda  r}\left(\gamma  \lambda ^4 (3 \alpha +\gamma  \Lambda )+3 Q^2\right)\,.
\ee
Note that for negative $\gamma=-\tilde \gamma$, the error function terms all continue to be real because
\be
\frac{\text{erfi}\left(\frac{\sqrt{\gamma } \lambda }{2 r}\right)}{\sqrt{\gamma } \lambda }
\longrightarrow
\frac{\text{erf}\left(\frac{\sqrt{\tilde \gamma } \lambda }{2 r}\right)}{\sqrt{\tilde \gamma } \lambda }\,.
\ee
At the asymptotic $r\rightarrow \infty$ region, the metric function $g_{tt}$ behaves
\be
-g_{tt}= - \ft13\Lambda r^2 - \ft16 \lambda^2 (3\alpha + 2\gamma\Lambda) - \fft{\mu}{r} +
\fft{6Q^2 - \Lambda \gamma^2\lambda^4}{24r^2} + \cdots\,.
\ee
Thus the solution is asymptotic to the AdS spacetime of cosmological constant $\Lambda$.
It is clear that the constant $\mu$ is associated with the condensate of the massless graviton.  The solution contains however additional slower falloff terms, which will not vanish owing to the no-ghost condition (\ref{noghost}).  The resulting ADM mass will then diverge, which is related to the divergence of the potential associated with the magnetic axion charges (\ref{nuint}). This problem persists even in Einstein gravity with $\gamma=0$, in which case the solution was obtained in \cite{Andrade:2013gsa}. In this paper, we adopt the concept of ``gravitational mass'' that is associated with the graviton mode \cite{Lu:2014maa}, namely
\be
M=\ft{1}{8\pi}\omega \mu\,.
\ee
As $r$ runs from the AdS boundary into the bulk, for sufficiently large $\mu$, there exists $r_0>0$ such that $\tilde f(r_0)=0$.  This gives rise to the event horizon of a black hole.
All the thermodynamic variables can be obtained, given by
\bea
T&=& \frac{e^{\frac{\gamma  \lambda ^2}{4 r_0^2}} \left(- 4 \Lambda  r_0^4- 2 \alpha  \lambda ^2 r_0^2-Q^2\right)}{16 \pi  r_0^3}\,,\qquad S=\ft14 \omega r_0^2 \big(1 - \fft{\gamma\lambda^2}{2r_0^2}\big)\,,\cr
\Phi &=& \frac{\sqrt{\pi } Q \text{erfi}\left(\frac{\sqrt{\gamma } \lambda }{2 r_0}\right)}{\sqrt{\gamma } \lambda }\,,\qquad Q_e =\fft{1}{16\pi}\omega Q\,,\qquad
P=-\fft{\Lambda}{8\pi}\,,\cr
V_{\rm th} &=& \ft12\omega \Big(2r_0 (2r_0^2 + \gamma\lambda^2) e^{\fft{\gamma\lambda^2}{4r_0^2}} - \sqrt{\pi} \gamma^2\lambda^4 \fft{\text{ erfi}(\fft{\sqrt{\gamma}\lambda}{2r_0})}{\sqrt{\gamma}\lambda}
\Big)\,.
\eea
As we discussed in section \ref{sec:thermal}, the entropy is non-negative provided that the no-ghost condition (\ref{noghost}) is satisfied.

    Note that if we follow the discussion at the end of section \ref{sec:thermal}, we can also introduce the axion magnetic charge $\lambda$ as the thermodynamic variables and obtain its conjugate potential.  The first law is then extended to (\ref{fl2}).  The formulae can be somewhat complicated and they do not add further clarity, and we therefore shall not present them here.

The above thermodynamic quantities are well defined for $\gamma>0$.  When $\gamma$ is negative, the situation can be somewhat subtle.  The ghost-free condition implies that $\gamma$ can be arbitrarily negative, and correspondingly the mass and volume defined above can hence be arbitrarily negative.  This is clearly unsatisfactory.  In order to resolve this issue, we introduce $\tilde \gamma=-\gamma>0$. We find that
\be
M \rightarrow -\fft{\omega}{96\sqrt{\pi}} \lambda^3 \sqrt{\tilde \gamma}(3\alpha -\Lambda \tilde \gamma)\qquad \hbox{as}\qquad \tilde\gamma \rightarrow +\infty\,.
\ee
This negative mass divergence can be resolved by shifting the coefficient $\mu$ to $\tilde \mu$ and define the mass as
\be
M=\ft{1}{8\pi}\omega \tilde \mu\,,\qquad
\tilde \mu = \mu -\ft{\sqrt{\pi}}{12}\Lambda (\tilde \gamma\lambda^2)^{\fft32} + \ft{\sqrt{\pi}}4
\alpha \lambda^3 \sqrt{\tilde \gamma}\,.
\ee
Consequently, the black hole volume for $\gamma=-\tilde \gamma<0$ is shifted by a
thermodynamic constant factor $\sqrt{\pi}\omega (\tilde \gamma\lambda^2)^{\fft32}$, namely
\be
V_{\rm th} = \ft12\omega \Big(2r_0 (2r_0^2 - \tilde \gamma\lambda^2) e^{-\fft{\tilde \gamma\lambda^2}{4r_0^2}} + \sqrt{\pi}\big( \tilde \gamma\lambda^2\big)^{\fft32} \big(1-\text{ erf}(\fft{\sqrt{\tilde \gamma}\lambda}{2r_0})\big)
\Big)\,.
\ee
(Recall that the parameters $(\lambda,\gamma)$ are not treated as thermodynamic variables in this paper.) It can be evaluated numerically that the volume is non-negative. Note that the volume formula for $\gamma>0$ and $\gamma<0$ are smoothly connected at $\gamma=0$.

Owing to the lack of the independent definition of the black hole mass and thermodynamic volume for our black holes, there is an ambiguity of adding extra constants to the volume formula. However, it appears to be rather {\it ad hoc} to add an extra term $\sqrt{\pi}\omega (-\gamma\lambda^2)^{\fft32}$ to formula as $\gamma$ crosses from positive to negative values, if it were to simply to avoid arbitrarily negative mass and volume. As we shall see in the next subsection, the situation becomes more severe in five dimensions. The volume formula satisfying the first law without analogous shift cannot be both real, making the shift mandatory. It is also worth noting that the shift quantity is purely ``numerical'' in that the quantity is not a thermodynamic variable and does not participate in the differentiations in the first law.  Finally, as we shall discuss in section \ref{sec:RII}, the ambiguity of the volume definition will not affect the key conclusion of the paper, namely the RII can be violated by the black holes in Horndeski gravity.

\subsection{Five dimensions}

As was discussed in section \ref{sec:adsansatz}, the functions $(h,a)$ were solved for general dimensions. In five dimensions, they are given by
\be
h = \sigma^{-2} f\,,\qquad a=\fft{2Q}{\gamma\lambda^2} \log\big(1 - \fft{\gamma\lambda^2}{4r^2}\big)\,,\qquad\sigma= 1 - \fft{\gamma\lambda^2}{4r^2}\,.
\ee
The function $f$ can be also solved easily from (\ref{hfeom}).  For $\gamma<0$, we have
\bea
f &=&-\ft1{12}\Lambda (2r^2 +\gamma\lambda^2) -\ft14\alpha\lambda^2 - \fft{\mu}{r^2} -
\fft{\gamma (3\alpha + \gamma \Lambda)\lambda^4}{48r^2} \log \big(1 - \fft{4r^2}{\gamma\lambda^2}\big)\cr
&& - \fft{Q^2}{3\gamma\lambda^2 r^2} \log\big(1-\fft{\gamma\lambda^2}{4r^2}\big)\,,\qquad
\hbox{with}\qquad \gamma<0\,.\label{d5sol1}
\eea
This solution is well-behaved for negative $\gamma$, which is consistent with the no-ghost condition (\ref{noghost}).  However, the no-ghost condition allows also a narrow window of
positive $\gamma$, in which case the function $f$ becomes complex at the asymptotic regions.  For positive $\gamma$, it is more natural to write
\be
\log\big(1 - \fft{4r^2}{\gamma\lambda^2}\big)=\log\big(\fft{4r^2}{\gamma\lambda^2}-1\big) + {\rm i}\pi\,.
\ee
The complex number can be absorbed into the redefinition of $\mu$ and the solution remains real at the asymptotic large $r$ region.  The solution becomes
\bea
f &=&-\ft1{12}\Lambda (2r^2 +\gamma\lambda^2) -\ft14\alpha\lambda^2 - \fft{\mu}{r^2} -
\fft{\gamma (3\alpha + \gamma \Lambda)\lambda^4}{48r^2} \log \big( \fft{4r^2}{\gamma\lambda^2}-1\big)\cr
&& - \fft{Q^2}{3\gamma\lambda^2 r^2} \log\big(1-\fft{\gamma\lambda^2}{4r^2}\big)\,,\qquad \hbox{with}\qquad \gamma>0\,.
\eea
Thus we see that in five dimensions, the reality condition requires us to use different forms of the solution to describe black holes for negative or positive $\gamma$.  This is analogous to the case in four dimensions, even though the reason is somewhat different.  Note that the $Q$-term and $\sigma$ take the same form as $\gamma$ crosses from negative to positive regions.  Note also that both the $\gamma>0$ and $\gamma <0$ solutions have the same limit of $\gamma\rightarrow 0$.

The thermodynamic quantities are all obtained in section \ref{sec:thermal}, except for $M$ and $V_{\rm th}$.  We find that for mass given by (\ref{genmass}), the black hole volume in five dimensions is given by
\bea
\gamma <0\,:&&\qquad V_{\rm th}=\ft1{4}\omega\Big( r_0^4+\ft12 \gamma  \lambda ^2 r_0^2 +\ft18\gamma ^2 \lambda ^4 \log \big(1-\frac{4 r_0^2}{\gamma  \lambda ^2}\big)
\Big)\,,\cr
\gamma>0\,:&&\qquad V_{\rm th}=\ft1{4}\omega\Big( r_0^4+\ft12 \gamma  \lambda ^2 r_0^2 +\ft18\gamma ^2 \lambda ^4 \log \big(\frac{4 r_0^2}{\gamma  \lambda ^2}-1\big)
\Big)\,.
\eea
Thus we see that the reality condition alone makes it mandatory to use different volume formulae for the $\gamma <0$ and $\gamma>0$ regions.  It is easy to check however that the two formulae are smoothly connected. In other words, both $V_{\rm th}$ and $dV_{\rm th}/d\gamma$ are continuous at $\gamma=0$.  Also note that the volume is non-negative in all the parameter regions of the black hole.

\section{Charged AdS planar black holes in general dimensions}

\label{sec:gend}

In the previous section, we discussed charged AdS planar black holes in four and five dimensions.  The different structures suggest that we should consider even or odd dimensions separately when we generalize to higher dimensions.

\subsection{Even dimensions}

We first consider the general even dimensions. We may solve (\ref{hfeom}) by
\be
f =  \sigma^{\fft{1}{D-4}} \tilde f\,,\qquad h=\sigma^{-\fft{1}{D-4}} \tilde f\,,
\ee
where $\sigma$ is given in (\ref{hfeom}) and $\tilde f$ is given in terms of hypergeometric functions
\bea
\tilde f &=& -\fft{2\Lambda r^2}{(D-1)(D-2)\sigma}\, _2F_1[-\ft12(D-1), \ft{1}{D-4},
-\ft12(D-3); \ft{(D-4)\gamma \lambda^2}{4r^2}]\cr
&& + \fft{Q^2}{2(D-2)(D-3)\sigma r^{2(D-3)}}\, _2F_1[\ft{1}{D-4}, \ft12(D-3);
\ft12(D-1), \ft{(D-4)\gamma\lambda^2}{4r^2}]\cr
&&-\fft{\alpha\lambda^2}{2(D-3)\sigma}\, _2F_1[-\ft12(D-3), \ft{1}{D-4}; -\ft12(D-5);
\ft{(D-4)\gamma\lambda^2}{4r^2}]\cr
&& - \fft{\mu}{\sigma r^{D-3}}\,.\label{tf}
\eea
The limit of this solution to $D=4$ is subtle and it is more advantageous to solve the equation (\ref{hfeom}) directly when $D=4$. For the large $r$, we have
\be
-g_{tt}=-\fft{2\Lambda}{(D-1)(D-2)} r^2 + c_0 + \fft{c_2}{r^2} + \fft{c_4}{r^4} + \cdots
+ \fft{c_{\fft{D-4}2}}{r^{D-4}} -\fft{\mu}{r^{D-3}} + \cdots\,,
\ee
where $c_i$ are constant functions of $(\alpha,\gamma,\lambda,\Lambda)$.  Thus solutions are all asymptotic to AdS.  Following the same argument in $D=4$, we consider gravitational mass defined by
\be
M=\fft{(D-2)\omega}{16\pi} \mu\,.\label{genmass}
\ee
For sufficiently large $M$, the solution can describe a black hole whose event horizon is located at $r_0$.  In section \ref{sec:thermal}, all the thermodynamic variables are derived except for mass and $V_{\rm th}$.  For the mass given by (\ref{genmass}), we find that the black hole volume is
\be
V_{\rm th}=  \fft{\omega}{D-1} r_0^{D-1}\, _2F_1[-\ft12(D-1), \ft{1}{D-4};
-\ft12 (D-3); \ft{(D-4)\gamma\lambda^2}{4r_0^2}]\,.\label{evenvth}
\ee
It can be easily verified that the first law (\ref{fl}) is satisfied.

The above solution is suitable to describe solutions with positive $\gamma$.  For negative $\gamma$, it suffers from the same problem as that in four dimensions, where the mass $M$ becomes divergent as $\gamma \rightarrow -\infty$, namely the mass becomes negative infinity in $D=4k$ dimensions and positive infinity in $D=4k+2$ dimensions.  For $\gamma<0$, we find that the form of the solution is better changed to the following
\bea
h &=& \sigma^{-\fft{2}{D-4}} f\,,\qquad \sigma= 1 - \fft{(D-4)\gamma\lambda^2}{4r^2}\,,\cr
f &=& \frac{8 \Lambda  r^4}{\gamma  (D-3) (D-2)^2 \lambda ^2}\, _2F_1[1,\ft12 (D+1);\ft12(D+1)+\ft{1}{D-4};\ft{4 r^2}{(D-4) \gamma  \lambda ^2}]\cr
&&+\frac{2 \alpha  r^2}{\gamma  \left(D^2-7 D+14\right)}\, _2F_1[1,\ft12(D-1);\ft12(D-1)+\ft{1}{D-4};\ft{4 r^2}{(D-4) \gamma  \lambda ^2}]\cr
&& + \frac{Q^2\sigma^{-\fft{D-5}{D-4}}}{2 (D-3) (D-2) r^{2(D-3)}} \, _2F_1[\ft{1}{D-4},\ft12(D-3);\ft12(D-1);\ft{(D-4) \gamma\lambda^2 }{4 r^2}]\cr
&& -\fft{\mu \sigma^{-\fft{D-5}{D-4}} }{r^{D-3}}\,.\label{oddsol}
\eea
This solution can be obtained from (\ref{tf}) by applying appropriately the hypergeometric function identity
\bea
_2F_1[a,b;c;z] &=&
\frac{\Gamma (c) \Gamma (b-a)}{\Gamma (b) \Gamma (c-a)}
\, (-z)^{-a}\, _2F_1[a,a-c+1;a-b+1;\ft{1}{z}]\cr
&&+\frac{\Gamma (c) \Gamma (a-b)}{\Gamma (a) \Gamma (c-b)}\, (-z)^{-b} \, _2F_1[b,b-c+1;-a+b+1;\ft{1}{z}]\,.
\eea
Note that the $Q$-term is kept the same as that in (\ref{tf}). It is easy to check that the hypergeometric functions are well behaved for negative $\gamma$, and the solution is also asymptotic to AdS.  The mass defined by (\ref{genmass}) is now non-divergent as $\gamma\rightarrow -\infty$. The thermodynamic volume is now given by
\be
V_{\rm th}=
-\frac{4 \omega  r^{D+1}\sigma_0^{\frac{D-5}{D-4}} }{\gamma  (D-3) (D-2) \lambda ^2}\, _2F_1[1,\ft12(D+1);\ft12(D+1)+\ft{1}{D-4};\ft{4 r_0^2}{(D-4) \gamma  \lambda ^2}]\,,
\label{oddvth}
\ee
where $\sigma_0$ is defined in (\ref{r0cons}) and $\gamma\in (-\infty, 0]$.

We now compare the solutions (\ref{oddsol}) with (\ref{tf}). The only difference between the two solutions is a constant shift of the mass parameter $\mu$ by
\be
\mu \rightarrow \mu + c_1 + \Lambda c_2\,,
\ee
where $c_1$ and $c_2$ are ``numerical'' constants.  By numerical, we mean that they depend only on the parameters $(\alpha, \gamma, \lambda)$ which do not involve in the thermodynamic variations.  This implies that the difference between the $V_{\rm th}$ defined in (\ref{evenvth}) and (\ref{oddvth}) must be numerical also.  Indeed, the difference is given by
\be
\fft{\omega \Gamma(\ft12(1-D))\Gamma(\ft12(D-1) + \ft{1}{D-4})}{2^D \Gamma(\ft{1}{D-4})}
\left(\fft{-1}{(D-4)\gamma\lambda^2}\right)^{\fft{1-D}{2}}\,.
\ee
This quantity is divergent for $\gamma\rightarrow -\infty$.  Since we do not have an independent way of calculating the mass and volume, there is an ambiguity of the definition of these quantities by shifting a ``numerical'' constant, and our analysis indicates that (\ref{evenvth}) and (\ref{oddvth}) are more suitable volume formulae for positive and negative $\gamma$ respectively.  However, as we shall see in section \ref{sec:RII}, this ambiguity does not affect the conclusion that the RII can be violated by black holes in Horndeski gravity.

In six dimensions, the hypergeometric functions reduce to ordinary functions and the volumes for positive and negative $\gamma$ are given by
\bea
\gamma>0: && V_{\rm th} = \ft{\omega}{15} r_0\big(3r_0^4 + 2\gamma\lambda^2 r_0^2 + 2\gamma^2\lambda^4\big) \sqrt{1-\ft{\gamma\lambda^2}{2r_0^2}}\,,\cr
\gamma=-\tilde\gamma<0: && V_{\rm th}=\ft{\omega}{15} r_0\big(3r_0^4 - 2\tilde\gamma\lambda^2 r_0^2 + 2\gamma^2\lambda^4\big) \sqrt{1+\ft{\tilde\gamma\lambda^2}{2r_0^2}} -
\ft{\sqrt2\,\omega}{15} (\tilde \gamma \lambda^2)^{\fft52}\,.
\eea

\subsection{Odd dimensions}

It turns out that when $D$ is odd, the hypergeometric functions in (\ref{tf}) all diverge except for the one associated with the $Q^2$. Thus for odd dimensions, the general expression of the solutions are given in (\ref{oddsol}), which are more suitable for the $\gamma<0$ case, and correspondingly the volume formula is given by (\ref{oddvth}). Indeed, when $D=5$, the solution reduces to (\ref{d5sol1}). In fact as in the five dimensional example, the solution (\ref{oddsol}) can also describe the case with $\gamma>0$, with some appropriate redefinition of $\mu$ to absorb the complex number so that the solution remains real.  This can be done for all odd dimensions. Since the $D=5$ example illustrates this point perfectly, we shall not elaborate it for general odd dimensions.

The solution degenerates in three dimensions, and we solve the case on its own. We find
\bea
f &=& \sigma^{-2} h\,,\qquad \sigma=1 + \fft{\gamma\lambda^2}{4r^2}\,,\qquad
a=Q \log \big(\frac{r}{\sqrt{\gamma } \lambda }\big)-\frac{\gamma  \lambda ^2 Q}{8 r^2}\,,\cr
h &=& -\Lambda r^2 - \ft12 (\alpha\lambda^2 + \gamma\lambda^2 \Lambda + Q^2)
\log \big(\fft{r}{\sqrt{\gamma}\lambda}\big) -\mu + \fft{\gamma\lambda^2 (\alpha\lambda^2 + Q^2)}{16r^2}\,.
\eea
The thermodynamic quantities are
\bea
M &=& \ft18\mu\,,\qquad T= \fft{-4\Lambda r_0^2 - \alpha\lambda^2 - Q^2}{8\pi r_0}\,,\qquad
S=\ft12\pi r_0 \big(1 - \fft{\gamma\lambda^2}{4r_0^2}\big)\,,\cr
Q_e &=& \ft18Q\,,\quad \Phi= \fft{\gamma\lambda^2 q}{8r_0^2} - Q\log(\fft{r_0}{\sqrt{\gamma}\lambda}\big)\,,\qquad P = -\fft{\Lambda}{8\pi}\,,\cr
V_{\rm th} &=& \ft12\pi \left(2 r_0^2 + \gamma\lambda^2 \log \big(\fft{r_0}{\sqrt{\gamma}\lambda}\big)\right)\,.
\eea
The solution describes the case with $\gamma>0$. Suitable care should be taken for negative $\gamma$ as well.

\section{Violations of RII}

\label{sec:RII}

The entropy and the volume of a Schwarzschild AdS black hole in general dimensions $D$ in Einstein gravity is given by
\be
S=\ft14 \omega \rho_0^{D-2}\,,\qquad V_{\rm th} = \fft{\omega}{D-1} \rho_0^{D-1}\,.\label{svth}
\ee
Interestingly, the volume of the Schwarzschild black hole is precisely equal to that of a spherical ball of radius $\rho_0$ in Euclidean space, namely
\be
V_{S} = \fft{\omega}{D-1} \rho_0^{D-1} = \fft{\omega}{D-1} \big(
\fft{4S}{\omega}\big)^{\fft{D-1}{D-2}}\,.\label{VE}
\ee
For a generic black hole with entropy $S$, we can define its effective radius $\rho_0$ by $S=\ft14 \omega \rho_0^{D-2}$.  For Einstein gravity with minimally coupled matter, $\rho_0$ is precisely the radius of the event horizon if the black hole is static. When the theory involve non-minimally coupled matter or higher order curvature invariants, $\rho_0$ may not be the radius of the event horizon.  We can nevertheless define the Euclidean volume (\ref{VE}).   The RII conjecture states that \cite{Cvetic:2010jb}
\be
V_{\rm th} \ge V_{S}\,.\label{rii1}
\ee
Schwarzschild AdS black hole in Einstein gravity saturates the inequality.  It was argued \cite{Feng:2017wvc} that this inequality in Einstein gravity with minimally coupled matter for static black holes is guaranteed by the null-energy condition.

    In higher derivative gravities, or gravities with non-minimally coupled matter, the entropy is no longer simply one quarter of the area of the horizon.  In this case, the RII conjecture expressed in (\ref{rii1}) is no longer purely geometrical.  There can thus be an alternative version of the RII conjecture, namely
\be
V_{\rm th} \ge V_{\cal A}\,,\qquad V_{\cal A} = \fft{\omega}{D-1}\, \big(\fft{{\cal A}}{\omega}\big)^{\fft{D-1}{D-2}}\,,\label{rii2}
\ee
where ${\cal A}$ is the area of the horizon.  This RII statement is purely geometrical.  The two statements (\ref{rii1}) and (\ref{rii2}) become equivalent in Einstein gravity with minimally-coupled matter.  They are asymptotically close to each other for large black holes when the Einstein-Hilbert term typically dominates.  We take the view that the version (\ref{rii1}) is more physically relevant, but we also study the version (\ref{rii2}). In fact, the inequality is saturated for static black holes in some ghost-free higher-derivative gravities such as Gauss-Bonnet and general Lovelock gravities.

     W now examine the inequality for the black holes of Horndeski gravity constructed in the previous sections.  First, as remarked at the beginning of section \ref{sec:adsansatz} that the RN black holes are solutions.  The entropy and volume for RN black holes are given by (\ref{svth}), and hence the RII is saturated, and the two statements of the RII are the same.

For all of our new black holes with non-vanishing axions, the situation is very different.
The corresponding Euclidean volumes, calculated from the entropy formula (\ref{genentropy}), take the form
\be
V_{S}= \fft{\omega}{D-1} r_0^{D-1} \Big(1 - \fft{(D-2)\gamma\lambda^2}{4r_0^2}\Big)
^{\fft{D-1}{D-2}}\,.
\ee
There is some ambiguity in our results for the thermodynamic volume up to some constant shifting of the ``numerical'' quantities such as $(\lambda, \gamma)$; however, for large enough horizon radius $r_0$, this ambiguity becomes insignificant.  We find that
\be
\fft{V_{\rm th}}{V_{S}} = 1  + \fft{(D-1)(D-2)\lambda^2}{4(D-3)r_0^2}\gamma + {\cal O}\big(\fft{1}{r_0^4}\big)\,.
\ee
The ambiguity of the definition of thermodynamic volume gives a contribution of order $1/r_0^{D-1}$ and hence can be ignored at large $r_0$.  Thus we see that for large $r_0$, the first statement of the RII conjecture holds for positive $\gamma$, and the saturation occurs when $\gamma=0$ or $r_0\rightarrow \infty$. On the other hand, for negative $\gamma$, the RII conjecture is violated.

    Analogously, for the second statement of RII, we find
\be
\fft{V_{\rm th}}{V_{\cal A}} = 1  + \fft{(D-1)\lambda^2}{4(D-3)r_0^2}\gamma + {\cal O}\big(\fft{1}{r_0^4}\big)\,.
\ee
Therefore, the dependence of the violation on the sign of $\gamma$ is the same as $V_{\rm th}/V_{S}$.

\begin{figure}[htp]
\begin{center}
\includegraphics[width=180pt]{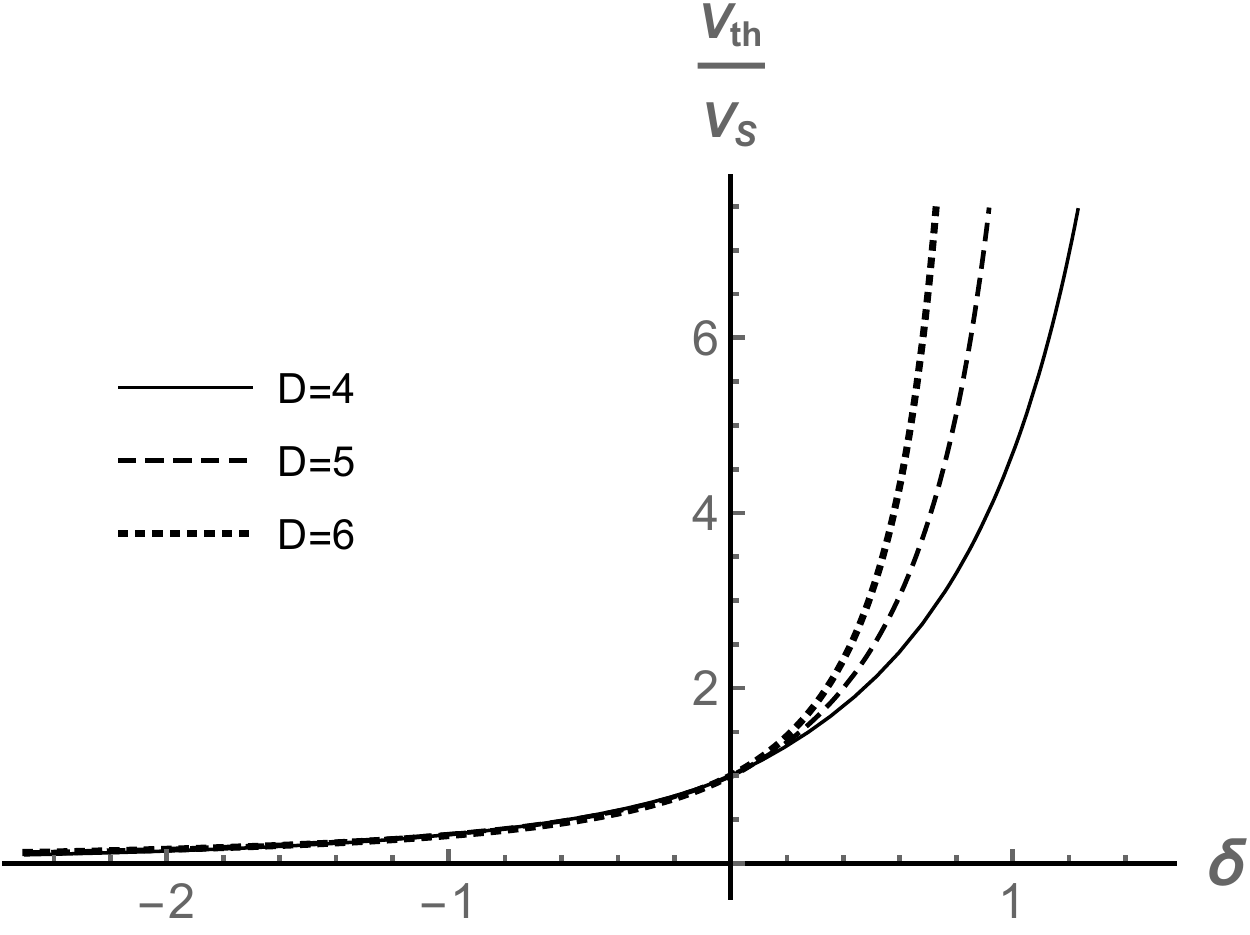}\qquad
\includegraphics[width=180pt]{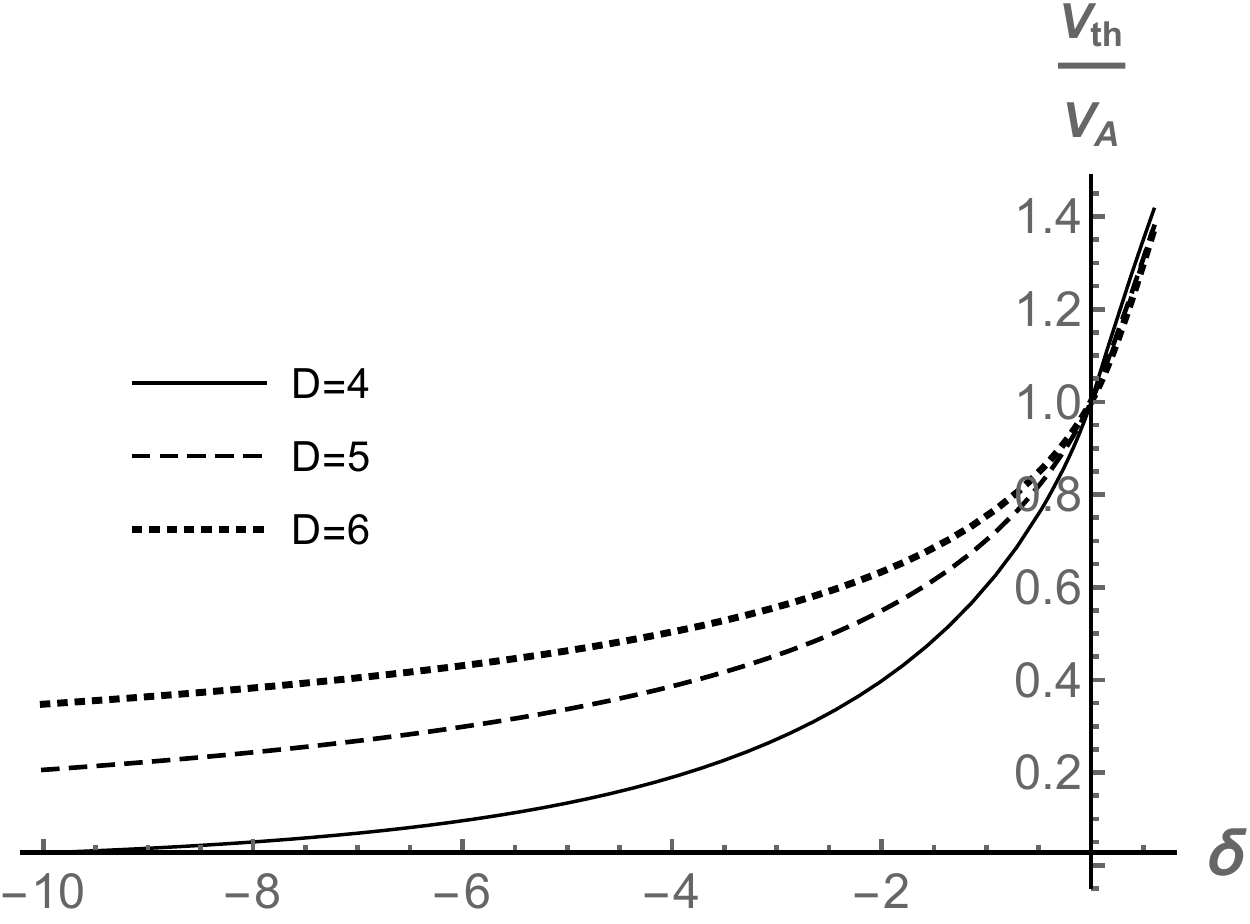}
\end{center}
\caption{\small The left plot gives the ratio $V_{\rm th}/V_{S}$ as a function of $\delta = \gamma\lambda^2/r_0^2$, and the right gives the ration $V_{\rm th}/V_{\cal A}$.  In both cases, the RII holds for $\delta>0$, saturated at $\delta=0$, and it is violated for $\delta <0$.  The ratio approaches zero as $\delta\rightarrow -\infty$.  The divergence of the left plot at the $\delta>0$ region implies that these black holes with finite volume carry no information, i.e~zero entropy.  On the other hand, as $\delta\rightarrow -\infty$, the black hole can carry an infinite amount of information in an infinitesimal volume.}
\label{figure1}
\end{figure}

It is of interest to plot the $V_{\rm th}/V_{S}$ and $V_{\rm th}/V_{\cal A}$ ratios as functions of $r_0$ and $\gamma$.  It turns out that for our determination of the volumes, this ratio depends only on the parameter ratio
\be
\delta = \gamma\fft{\lambda^2}{r_0^2}\,.
\ee
As concrete examples, we give the ratio explicitly in four, five and six dimensions:
\bea
D=4:&& \fft{V_{\rm th}}{V_{S}} =
\left\{
  \begin{array}{ll}
    \frac{2 e^{\delta /4} (\delta +2)-\sqrt{\pi } \delta ^{3/2} \text{erfi}\left(\frac{\sqrt{\delta }}{2}\right)}{\sqrt{2} (2-\delta )^{3/2}}\,, &
\qquad \delta\ge 0\,, \\
\frac{2 e^{\delta /4} (\delta +2)+\sqrt{\pi } (-\delta) ^{3/2}\left(1- \text{erf}(\frac{\sqrt{-\delta }}{2})\right)}{\sqrt{2} (2-\delta )^{3/2}}
\,, &\qquad \delta\le 0\,;
  \end{array}
\right.\nn\\
D=5:&& \fft{V_{\rm th}}{V_{S}}=
\left\{
  \begin{array}{ll}
  \frac{\delta ^2 \log(\frac{4}{\delta }-1)+4 \delta +8}{2^{1/3} (4-3 \delta )^{4/3}}\,, &\qquad\delta\ge 0\,, \\
\frac{\delta ^2 \log(1-\fft{4}{\delta})+4 \delta +8}{2^{1/3} (4-3 \delta )^{4/3}}\,, &\qquad \delta\le 0\,;
  \end{array}
\right.\nn\\
D=6:&& \fft{V_{\rm th}}{V_{S}}=
\left\{
  \begin{array}{ll}
\frac{\sqrt{2-\delta } (2 \delta ^2+2 \delta +3)}{3 \sqrt{2} (1-\delta )^{5/4}}\,, & \qquad\delta\ge 0\,, \\
\frac{\sqrt{2-\delta }(2 \delta ^2+2 \delta +3) - 2(-\delta)^{\fft52}}{3 \sqrt{2} (1-\delta )^{5/4}}\,, &\qquad\delta\le 0\,.
  \end{array}
\right.
\eea
The three functions and their derivatives are continuous at $\gamma=0$.  The function $V_{\rm th}/V_{\cal A}$ is given by
\be
\fft{V_{\rm th}}{V_{\cal A}}= \fft{V_{\rm th}}{V_S} \big(1 - \ft14(D-2)\delta\big)^{\fft{D-1}{D-2}}\,.
\ee
The plots of $V_{\rm th}/V_{S}$ and $V_{\rm th}/V_{\cal A}$ as functions of $\delta$ in $D=4,5,6$ are presented in Fig.~\ref{figure1}.  The divergence of the left plot in the $\delta>0$ region implies that the black hole is not efficient for storing information. For a given horizon radius $r_0$, it has the finite thermodynamic volume, but stores no information (zero entropy).  On the other hand, as $\delta\rightarrow -\infty$, the entropy becomes infinite, whilst the thermodynamic volume shrinks to zero.

\section{Conclusions and discussions}

In this paper, we studied Einstein-Horndeski-Maxwell gravity with a cosmological constant and multiple Horndeski axions.  We constructed a new class of AdS planar black holes where the axions span over the planar directions.  We study the black hole thermodynamics and derive the black hole volume by means of completing the first law, in which the cosmological constant is treated as a thermodynamic variable, the pressure.  By comparing the entropy and the volume, we find that the RII is violated when the Horndeski coupling constant $\gamma$ is negative.  Owing to the fact that in Einstein-Horndeski-Maxwell gravity, the entropy of a black hole is no longer simply one quarter of the area of the horizon, we presented two RII statements, one based on the entropy and the other based on the area of the horizon, and we found that both RII's could be violated.

As we discussed in section \ref{sec:horndeski}, the ghost-free condition of the black hole backgrounds allows that $\gamma\rightarrow -\infty$. In this limit, the entropy (\ref{genentropy}) becomes infinite, even when the area of the horizon remains finite. It follows from (\ref{VE}) and (\ref{rii2}) that we have
\be
\fft{V_S}{V_{\cal A}}=\big(1 - \ft14 (D-2)\delta\big)^{\fft{D-1}{D-2}}\rightarrow +\infty\,,\qquad
\hbox{as}\qquad \delta\rightarrow -\infty\,.
\ee
The violation of the RII aside, the ability of storing arbitrarily large amount of information on a finite physical horizon area is counter intuitive.  This phenomenon occurs also in Einstein-Gauss-Bonnet gravity whose black hole entropy (\ref{gbsv}) becomes divergent as $\alpha\rightarrow +\infty$. However, in the Gauss-Bonnet theory, the quantum effects imply that the causality will be violated when the coupling constant $\alpha$ is above some critical value and the classical Lagrangian cannot be trusted \cite{Hofman:2008ar}.  The similar $S/\gamma$ dependence (\ref{genentropy}) suggests that a similar causality bound may exist in Horndeski gravity which may put an up bound on how much entropy that can be stored in a given volume.

It is of interest to note that although we obtain the volume formulae by indirect means of completing the first law in the case by case basis, there is a general local expression for the volumes of all our solutions.  To see this, note that we can write
\be
-g_{tt} = -\fft{2\Lambda}{(D-1)(D-2)} \tilde h(r) + h_0(r)\,,
\ee
where $(\tilde h,h_0)$ do not involve any $\Lambda$.  The thermodynamic volumes for the black holes we constructed in this paper can all be expressed locally as
\be
V_{\rm th}=\fft{\omega}{D-1} r_0^{D-3}\,\tilde h(r_0)\, \sigma_0^{\fft{D-3}{D-4}}\,.
\ee
This formula is analogous to the one obtained in \cite{Feng:2017wvc} for black holes in two-derivative gravities.

The new black holes we constructed generalize the four dimensional one \cite{Jiang:2017imk} for which the holographic transport properties with momentum dissipation by the Horndeski axions were studied \cite{Jiang:2017imk,Baggioli:2017ojd}.  In addition to our focus on studying black hole volumes, we expect these solutions in general dimensions can also have the similar applications.  In particular, before ending the paper, we would like to comment on the butterfly velocity that was introduced in \cite{Shenker:2013pqa}. For our black holes, it is simply given by
\be
v_{\rm B}^2 = \fft{(D-1) T}{8 \sigma_0^{1/(D-4)}\,r_0\,P} =\fft12 \fft{TS}{V_{\cal A} P}\, \sigma_0^{-1/(D-4)} \Big(1 - \fft{(D-2)\gamma \lambda^2}{4r_0^2}\Big)^{-1}\,,
\ee
where $T,S$, $P$ and $\sigma_0$ were given in section \ref{sec:horndeski} and $V_{\cal A}$ is defined in (\ref{rii2}). (The four dimensional result was given in \cite{Baggioli:2017ojd}, although it was not put in this form.)  In Einstein gravity, surveyed for a variety of isotropic AdS planar black holes, the butterfly velocity can be expressed as \cite{Feng:2017wvc}
\be
v_{\rm B}^2 = \fft12\fft{TS}{V_{\rm th} P}\,.
\ee
We find that this formula is no longer true in general Horndeski gravity, but instead we have the inequality
\be
v_{\rm B}^2\quad\rightarrow\quad
\left\{
  \begin{array}{ll}
     >\fft12\fft{TS}{V_{\rm th} P} , & \qquad \gamma > 0 \\
     < \fft12\fft{TS}{V_{\rm th} P}, & \qquad \gamma <0
  \end{array}
\right.
\ee
The equality occurs at $\gamma=0$ or in the large black hole $r_0\rightarrow \infty$ limit.

\section*{Acknowledgement}

The research of X.-H.F, W.-T.L and H.L.~is supported in part by NSFC grants No.~11175269, 11235003 and 11475024. H-S.L.~is supported in part by NSFC grants No.~11305140, 11375153,
11475148 and 11675144.

\end{document}